\documentclass[twocolumn,superscriptaddress,amsmath,amssymb,floatfix,aps,notitlepage,nokeys,pre]{revtex4}
\usepackage{graphicx}
\usepackage{dcolumn}
\usepackage{bm}
\usepackage{xcolor}
\usepackage{hyperref}

\newcommand{\sg}[1]{}
\renewcommand{\sg}[1]{{\color{red}{#1}}} 

\begin{document}

\title{Giant and Persistent Spatial Amplification from FB in a Continuum Elastic Lattice}
\title{Giant Response Amplification via FB in Lattices with Inertial Anchors}
\title{Giant and Persistent Response Amplification from FB via Inertial Anchors}
\title{Giant Response Amplification from FB via Inertial Anchors}
\title{Giant Flat Band Response Amplification via Inertial Anchors}
\title{Giant Flat Band Amplification via Inertial Anchors}

\author{Wentao Mao}
\affiliation{Department of Civil, Environmental, and Geo- Engineering, University of Minnesota, Minneapolis, Minnesota, USA}

\author{Stefano Gonella}
\email{sgonella@umn.edu}
\affiliation{Department of Civil, Environmental, and Geo- Engineering, University of Minnesota, Minneapolis, Minnesota, USA}

\date{\today}

\begin{abstract}
	In electronic materials, flat bands are associated with compact electron localization, with implications for superconductivity, ferromagnetism and strongly correlated systems. The physical significance of their counterparts in elastic media is far less charted. Here we report a strategy to achieve elastic flat bands through an inertial retrofitting of classical lattice architectures. The idea is to alter the cell geometry to realize a network of inertial anchors, effectively partitioning the lattice into an array of weakly coupled emergent resonators, whose resonances appear as flat bands in the phonon spectrum. 
    We demonstrate flat-band conditions that combine localized and extended state attributes and induce giant 
    response amplification that is spatially and temporally persistent. Laser vibrometry experiments reveal three signatures of this mechanism: amplification up to two orders of magnitude compared to pass band and band gap conditions, multi-cell activation that is agnostic to the source location, and a persistent transient response even after several excitation cycles. 
\end{abstract}

\maketitle

Controlling the localization and storage of energy is a central theme in engineered lattices and metamaterials.
Several mechanisms can arrest wave transport, including defects ~\cite{deng2022observation,jo2022revealing}, flat bands (FB) ~\cite{shen2022observing,riva2025creating}, and topologically protected boundary, interface or corner modes~\cite{Paulose-et-al_Topological-Dislocations_Nature_2015,Makwana-Craster_Topological-Network_PRB_2018,fan2019elastic,Serra-Gracia-et-al-Quadrupole-topo-insulator_Nature_2018,azizi2023dynamics,azizi2024omnidirectional}.
In this panorama, FB are distinct because they are spectral features of the bulk rather than states tied to a prescribed defect, boundary, or interface.
A FB is a branch with nearly vanishing group velocity and a high density of states, allowing energy to remain confined near the excitation~\cite{karki2023non,samak2024direct}.
Flat-band physics was first explored in electronic materials ~\cite{sutherland1986localization,tasaki1994stability,maksymenko2012flat,misumi2017new,chase2024compact, kang2020topological,sun2011nearly,wan2023topological,sarkar2025symmetry} and have since been extended to electromagnetic~\cite{leykam2013flat,mukherjee2015observation,travkin2017compact,song2023multiple}  and acoustic systems~\cite{zheng2014acoustic,dubois2019acoustic,tang2024systematic,karki2023non,shen2022observing,samak2024direct,riva2025creating,ling2026twist,han2025all,zhang2025observation,ramos2025flat}.
In acoustics, FB have been shown to enable compact localized states and response amplification in structures such as coupled resonator lattices~\cite{samak2024direct}, kagome-type lattices~\cite{karki2023non}, twisted Moiré lattices~\cite{ling2026twist}, and decorated waveguide networks~\cite{riva2025creating}.

In contrast, the occurrence,  physical significance and technological opportunities of FB for elastic media remain largely unexplored.
Recent developments have laid down platforms to realize FB in mechanical systems and have shown that flat-band excitation can lead to strong energy localization. A key development came from the study in~\cite{samak2024direct}, which demonstrated achievement of all-flat-band conditions and strong vibration localization in discrete particle network systems. Other studies have invoked destructive-interfering mechanisms between neighboring elements to achieve non-singular FB in elastic chains endowed with ground stiffness~\cite{Riva-et-al_Nonsingular-Flat-Band_JSV_2025}. 
Previously, emergence of FB had been demonstrated, among other effects, in lattices of weakly coupled solid patches~\cite{Matlack-et-al_Discrete-Models_Nat-Mat_2018}.

In most existing realizations, the amplitude-boosting effects of FB are achieved by exploiting some type of destructive interference and are concentrated at a small number of sites surrounding the excitation point.
For applications to vibration control, however, other secondary factors beyond localization alone are highly desirable and need to be targeted simultaneously.
In vibration-based harvesting, the crucial requirement is to localize the response and maximize its amplitude to elicit large strains that can be converted to appreciable electric outputs; other important requirements include that the strain activation be \textit{spatially pervasive}, \textit{source-agnostic} and \textit{temporally persistent}, in order to harvest with comparable efficiency from the entire vibrating domain, independent of the location of the excitation and over prolonged time windows. These considerations raise the following question: can we engineer FB conditions that combine localized and extended state attributes, yielding large-amplitude response that is strongly localized within each cell and yet delocalized over multiple adjacent cells?

In this Letter, we answer this question by reporting an intuitive strategy for the realization of elastic FB that only requires an inertial ``retrofitting" of classical lattice architectures. Importantly, our approach replaces canonical interference methods with the establishment of a network of weakly coupled resonators as the FB-enabling mechanism. 
As a result, we are able to demonstrate strong response-boosting signatures, including two-order-of-magnitude displacement amplification and persistent transient response, as well as the additional attribute of a spatially pervasive multi-cell activation.

\begin{figure*}[t]
	\centering
	\includegraphics[width=0.95\textwidth]{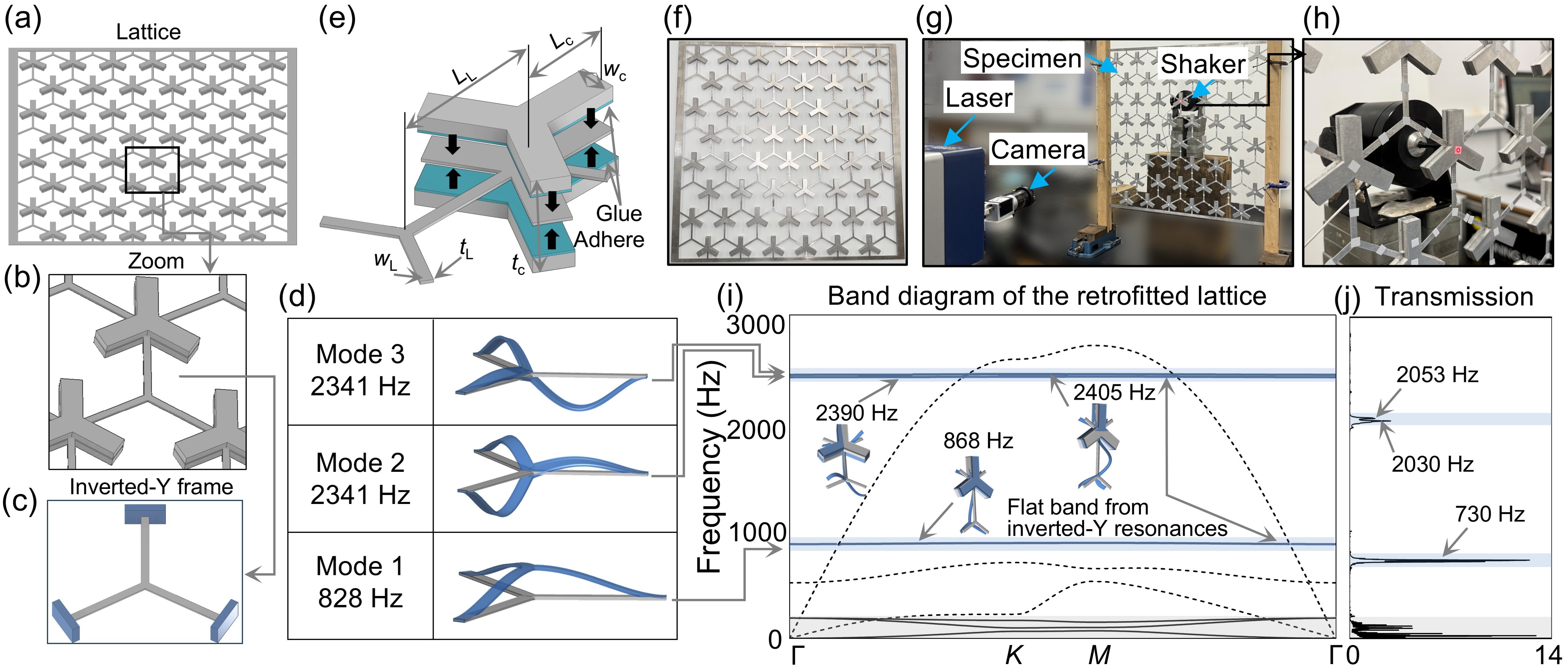}
	\caption{\label{fig:mechanism}
		Strategy for realizing elastic flat bands by retrofitting a hexagonal
        lattice with inertial anchors: Design, simulations and prototype. (a) Lattice configuration with inertial anchors.
        (b) Details of anchors highlighting OOP thickness contrast with primary lattice.
        (c) Emergent inverted-Y resonator with virtual clamps.
        (d) Calculated natural frequencies and mode shapes of inverted-Y frame.
		(e) Geometry details and multi-layer fabrication strategy.
        (f) Water-jet cut aluminum specimen.
		(g) Experimental setup.
        (h) Detail of actuation.
		(i) Lattice band diagram, with emerging flat bands highlighted as blue lines. 
		(j) Measured transmission spectrum of the lattice, revealing signatures of flat bands.
	}
\end{figure*}

The idea is to augment a baseline lattice configuration by inserting at selected sites arrays of so-called ``inertial anchors", i.e., auxiliary (non connectivity-changing) 
elements. The purpose of the anchors is to inertially constrain the sites to which they are attached and suppress their mobility, thereby achieving by inertia a grounding effect analogous to an elastic foundation. When the anchors are spatially arranged so as to confine and dynamically isolate selected regions of the lattice, these regions behave, near their resonant frequencies, as weakly coupled emergent resonators. In this frequency range, the anchored sites remain nearly stationary and act as structural nodes, allowing motion trapping within each resonator while only weakly transmitting across adjacent resonators. Each resonator exhibits a spectrum of resonances that are expected to manifest in the lattice phonon band diagram as FB. Interestingly, and in contrast with common FB activation strategies, this approach is frequency-dependent: outside the FB range, the anchoring effect is diluted, inter-cell coupling is restored, and the lattice can still support pass band (PB) modes.

\begin{figure*}[!t]
	\centering
	\includegraphics[width=1\textwidth]{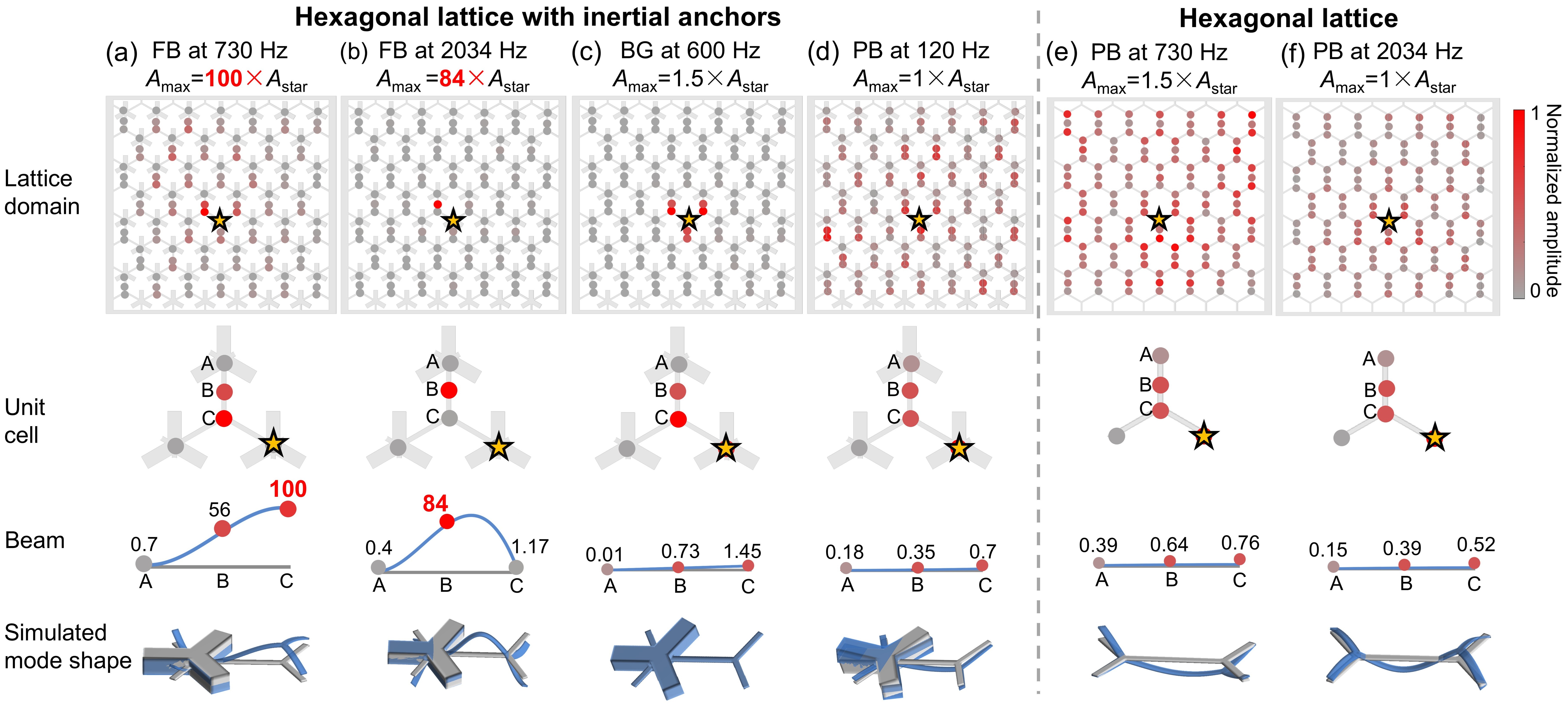}
	\caption{\label{fig:amplification}
		Experimental evidence of flat-band-induced amplitude amplification in lattice with anchors.
		(a-d) OOP displacements acquired via laser vibrometry for excitations at FB (a-b), band gap (BG) (c), and pass band (PB) (d) frequencies. For each case we show the wave field data over the entire lattice domain, displacements at scan points within a representative cell, and reconstructed displacement profile along a beam of the primary lattice, matching well the corresponding calculated mode shape.
		The maximum value in red is the largest measured displacement normalized by the displacement at the excitation point.
		The first FB (a) produces a maximum amplification of $100\times$, with strong deformation concentrated in the inverted-Y frame.
		Band-gap excitation in (c) shows weak and localized response. Pass-band excitation in (d) shows extended response but no significant amplification.
		(e,f) Regular hexagonal lattice excited at the same frequencies as in (a) and (b), showing standard pass band conditions without the amplitude boosting granted by resonance activation.
	}
\end{figure*}

We seek a blueprint to realize inertial anchors by re-purposing a cell-decoration technique we had originally introduced for locally-resonant BG~\cite{gonella2009interplay,Celli-et-al_Culring_SMS_2017}. We start from a thin hexagonal lattice and we attach at each site three cantilever beams directed along the bisectrices of the bond angles [Fig.~\ref{fig:mechanism}(a)]. 
We pick a large out-of-plane (OOP) thickness contrast between the primary lattice and the cantilevers 
[Fig.~\ref{fig:mechanism}(b)]. Thus, under OOP excitation, the cantilever-endowed sites feel the large inertia and their motion is penalized as if anchored to a foundation.  
As a result, the region bounded by the anchors, which is shaped like an \textit{inverted Y}, can be treated as a three-legged frame clamped at its edges and behaves as an emergent resonator [Fig.~\ref{fig:mechanism}(c)]. The lattice effectively morphs into an array of resonators that are only loosely coupled due to the imperfect nature of the clamps. 

We study the resonator dynamics via finite element analysis (FEA), assuming the following parameters: 
lattice beam length, OOP thickness and in-plane (IP) width $L_{\mathrm{L}}=50~\mathrm{mm}$, $t_{\mathrm{L}}=1~\mathrm{mm}$, $w_{\mathrm{L}}=4~\mathrm{mm}$, cantilever length, OOP thickness and IP width $L_{\mathrm{c}}=30~\mathrm{mm}$, $t_{\mathrm{c}}=10.6~\mathrm{mm}$, $w_{\mathrm{c}}=12~\mathrm{mm}$;
for Aluminum (Al), Young's modulus $E=69~\mathrm{GPa}$, Poisson's ratio $\nu=0.33$, and density $\rho=3000~\mathrm{kg/m^3}$.
The three lowest natural frequencies and mode shapes of the inverted-Y frame are reported 
in Fig.~\ref{fig:mechanism}(d). Note that, because of the approximate symmetry of the frame, the second and third modes are nearly degenerate.
The lattice band diagram, computed via standard FEA-enabled Bloch analysis~\cite{Phani-et-al_Periodic_JASA_2006}, is shown in Fig.~\ref{fig:mechanism}(i). In addition to propagating OOP (solid lines) and IP (dashed) modes, we observe the emergence of three OOP FB (blue solid lines). 
Their frequencies and mode shapes match those of the inverted-Y frame, confirming that the FB originate from emergent resonators. 
A more formal rationale to link FB and emergent resonators can be obtained by invoking an analogy with a tight binding model of a network of coupled resonators in the limit of vanishing strength of coupling (see SI for details) ~\cite{brillouin1953wave, yariv1999coupled, longhi2015tunable, nfor2023modulational}. It can be shown that the dispersion relation for a weak chain of resonators reduces to the form $\omega(k) \approx \omega_0+2\kappa\cos(ka)$, 
where $\omega$, $k$ and $a$ are frequency, wave number and lattice constant, and $\kappa$ is the coupling strength between resonators. Clearly, when $\kappa$ is finite, the isolated resonance frequency $\omega_0$ broadens into a narrow band
with bandwidth $\Delta \omega = 4|\kappa|$. 
Here, when the anchors are active, the coupling between inverted-Y resonators becomes weak, i.e.,  $|\kappa| \ll \omega_0$, 
$\Delta\omega$ vanishes and the band becomes nearly dispersionless: $\omega(k) \approx \omega_0$.

We test these arguments on a physical prototype. To reduce damping we work with a metallic lattice. The challenge is to realize a specimen with variable OOP thickness without resorting to 3D printing, which is prohibitive for metal parts at our scale of interest. For this, we devise a two-stage fabrication strategy that involves only cutting. We first fabricate a \textit{uniform-thickness} specimen 
(including cantilevers) by water-jet cutting a 1-mm thick Al slab [Fig.~\ref{fig:mechanism}(f)]. We then cut 4.8 mm-thick platelets shaped as triads of cantilevers and glue them 
to the appropriate sites as illustrated in Fig.~\ref{fig:mechanism}(e). 
We excite OOP via a shaker (Brüel $\&$ Kjær Type 4810 Mini) and we measure the OOP response using a scanning laser Doppler vibrometer (Polytec PSV-400) [Fig.~\ref{fig:mechanism}(g-h)].
The transmission is calculated by averaging the displacement over a ring of scan points surrounding the excited cell and normalizing it by the displacement at the excitation point (details in SI).
Figure~\ref{fig:mechanism}(j) shows the transmission spectrum under 
pseudorandom excitation. In addition to the activation of low-frequency PB, we record strong transmission peaks at two frequencies above the OOP bulk band, which we recognize as signatures of FB. 
The peaks are shifted slightly downward compared to numerical predictions. This is attributed to the imperfect bonding between the layers, which produces non-ideal clamping conditions, 
increasing the effective length of the inverted-Y beams and lowering their resonances. 

\begin{figure*}[!t]
    \centering
    \includegraphics[width=0.85\textwidth]{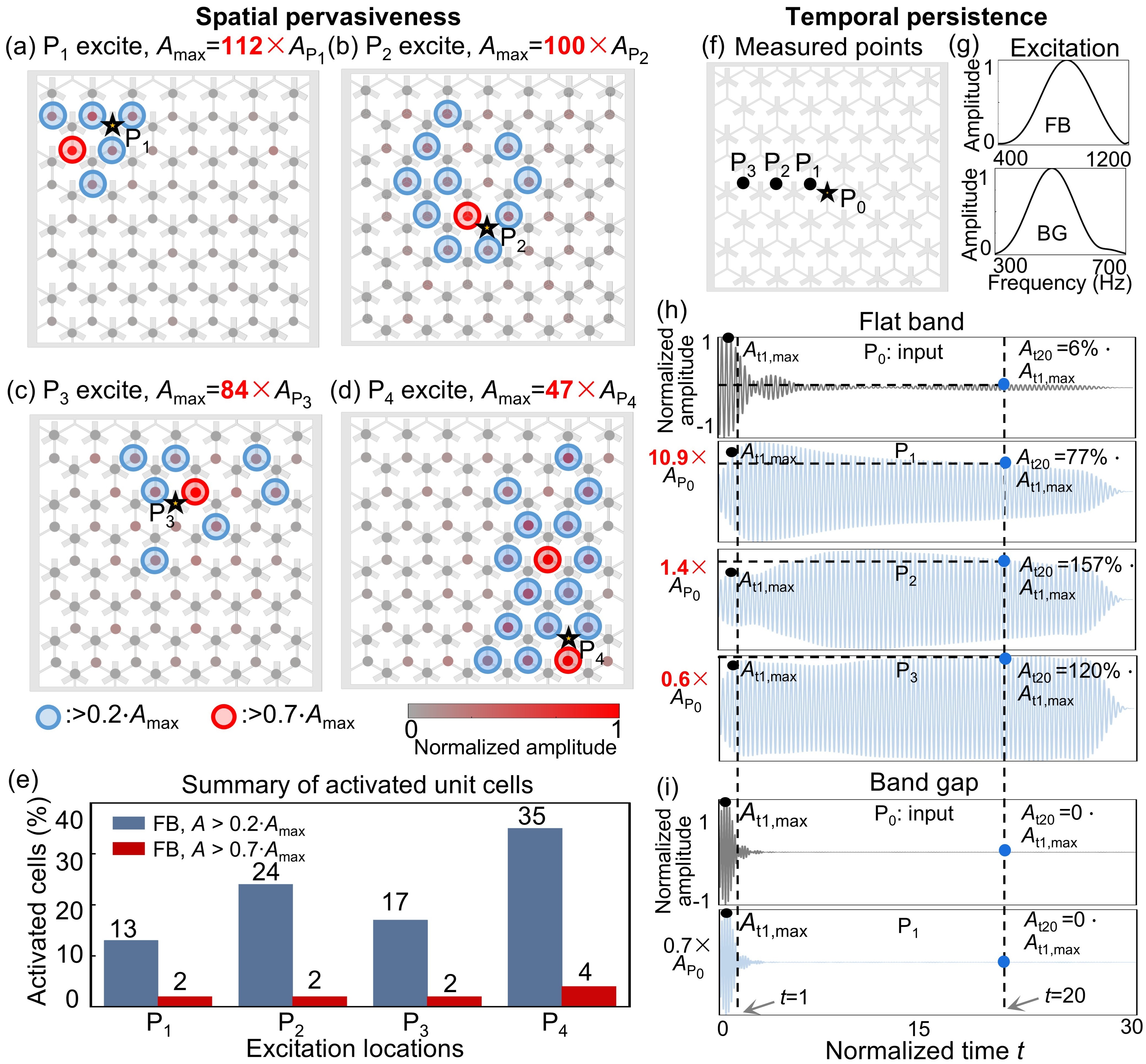}
    \caption{\label{fig:spatial and temporal}
    Experimental evidence of spatial pervasiveness and temporal persistence of flat band activation. (a--d) Spatial response maps at the FB frequency for four excitation locations (marked by stars). Strong multi-cell activation across cases indicates that the effect is spatially pervasive and largely agnostic to source location.
    Blue and red circles denote cells whose normalized amplitudes exceed 20\% and 70\% of the maximum amplitude, respectively.
    (e) Fractions of activated and highly activated cells at the flat-band and band-gap frequencies, showing consistency of activation.
    (f-g) Scan points for transient measurements and spectra of chirps used as excitations.
    (h-i) Transient responses recorded in FB (h) and BG (i) conditions. Time $t$ is normalized by the excitation duration $T_{\mathrm{exc}}$, such that $t=1$ corresponds to the end of the three-cycle chirp. For FB, the amplitude boosting effect lingers, with amplitude preserved even after long periods (at a time corresponding to 20 excitation periods).
    }
\end{figure*}

To quantify the amplitude amplification, we perform scans of the entire lattice (three scan points per cell: two lattice sites and one mid-beam point). 
We compare representative frequencies in the FB, BG, and PB ranges [Fig.~\ref{fig:amplification}(a-d)]. For each case, we show the OOP response field over the entire lattice and within a cell, and the profile along one of the frame beams. At the FB frequencies [Fig.~\ref{fig:amplification}(a-b)], we record large displacement amplitudes with peaks at the centers of the inverted-Y resonators. As expected, the sites with cantilevers do not displace appreciably. 
We introduce the amplification metric $A_{\mathrm{max}}/A_{\mathrm{star}}$, defined as the ratio between the maximum amplitude measured at a scan point and that at the excitation point. 
For the first FB, the amplification reaches $100\times$. 
The amplitude maps also reveal that, because the FB has a finite but narrow bandwidth, weak inter-cell coupling allows the excitation to access other resonant frames, leading to a \textit{spatially pervasive} effect characterized by high-amplitude response over multiple cells. 
In contrast, BG excitation [Fig.~\ref{fig:amplification}(c)] suppresses transmission and confines the response near the excitation point. PB excitation [Fig.~\ref{fig:amplification}(d)], on the other hand, promotes propagation through the lattice, achieving comparable spatial pervasiveness, but does not produce appreciable local amplification.
For reference, we also test a pristine hexagonal lattice (with identical properties) featuring PB at the FB frequencies of the lattice with anchors [Fig.~\ref{fig:amplification}(e-f)]. The band diagram and measured transmission spectrum can be found in the SI. 
Here, too, we observe full transmission that effectively insonifies the entire domain but lacks amplification, with the response remaining on the order of $A_{\mathrm{star}}$.

To further test the spatially pervasive attribute, we excite the lattice at several locations $\textrm{P}_{\textrm{i}=1 \dots 4}$ near the first FB frequency 
[Fig.~\ref{fig:spatial and temporal}(a-d)]. 
We report large peak amplitudes in all cases, with amplification levels (defined as maximum recorded amplitude $A_{\mathrm{max}}$ normalized by the actuation point amplitude $A_{\mathrm{P}i}$) ranging between $47 \times$ and $112 \times$. Next, we count the number of cells where the response boosting exceeds prescribed thresholds.
Cells with amplification levels $>$ 20\% and 70\% of $A_{\mathrm{max}}$  
are classified as \textit{activated} and \textit{highly activated}, respectively. The activated cell fraction ranges from 13\% to 35\%, with some highly activated cells observed in every case [Fig.~\ref{fig:spatial and temporal}(e)]. This analysis reveals that the amplification effect is robust against changes in the excitation location, suggesting that the approach is largely \textit{source agnostic}. Interestingly, by varying the thickness contrast between anchors and lattice, we can control the degree of flatness of the FB, which reflects the strength of inter-cell coupling, obtaining effectively a nob to tune between 
localization and spread of the response (see parametric study in SI).

\begin{figure}[!t]
    \centering
    \includegraphics[width=1\columnwidth]{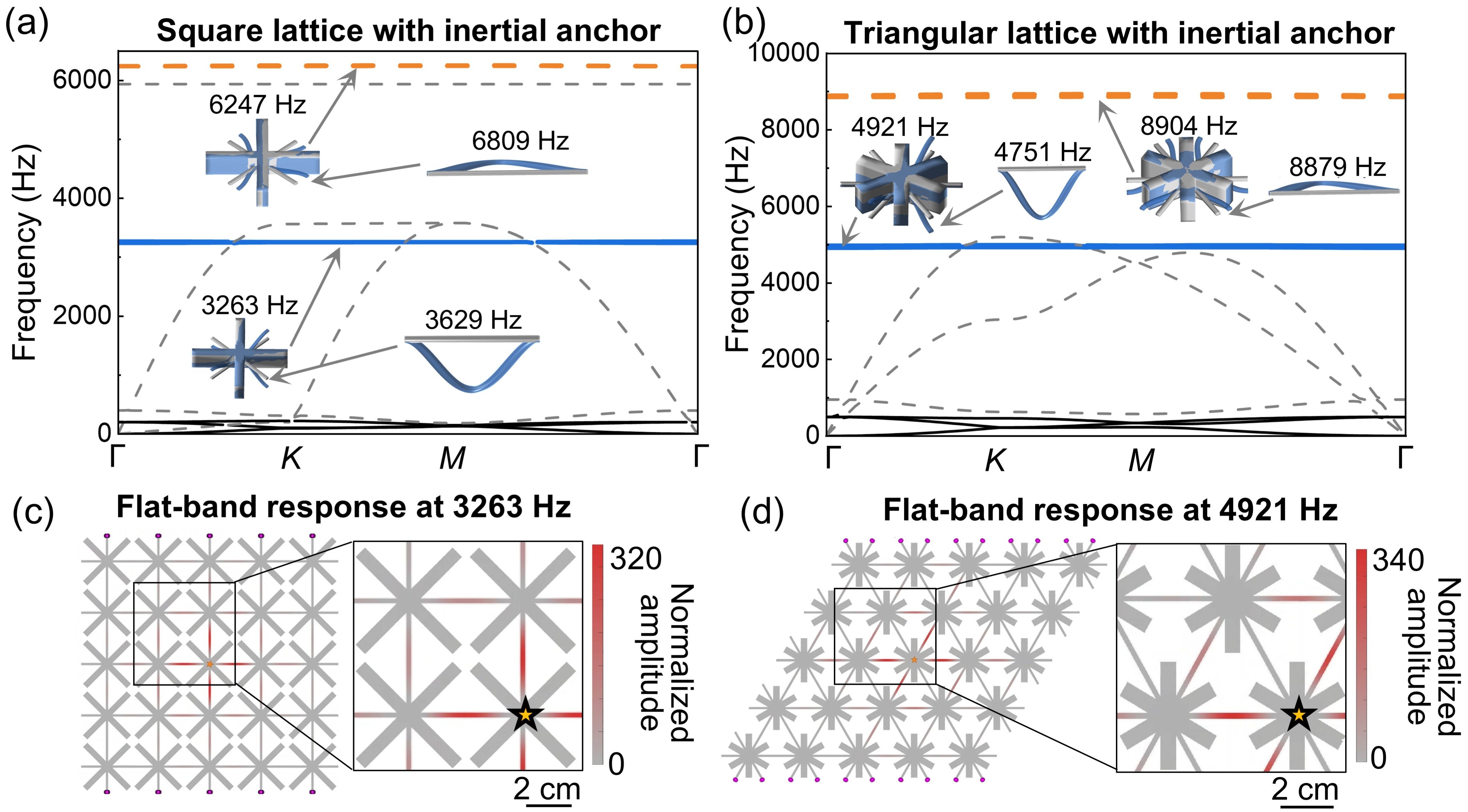}
    \caption{\label{fig:generalization}
    Versatility of inertial anchor approach for FB via examples of square and triangular lattices.
    (a),(b) Band diagrams of lattices retrofitted with inertial anchors. The blue and orange curves highlight two nearly flat bands.
    The insets show representative mode shapes of FBs and the corresponding connecting-beam resonances.
    (c),(d) Spatial responses of 5×5 square and triangular lattice domains at the first flat-band frequencies, $3263~\mathrm{Hz}$ and $4921~\mathrm{Hz}$, respectively. The yellow star marks the excitation point, while the pink points along the upper and lower boundaries indicate the constrained locations. The displacement amplitude in each lattice is normalized by the amplitude at the excitation point.
    }
\end{figure}

The same weak-coupling mechanism also gives rise to persistent transient dynamics. To show this, we excite the structure with a three-cycle chirp containing the flat-band frequency (spectrum in [Fig.~\ref{fig:spatial and temporal}(g)]) and track the displacement envelope during the excitation phase ($0\!<\!t\!<\!1$) and the relaxation period ($1\!<\!t\!<\!30$). We compare three sample points $\textrm{P}_{\textrm{i}=1 \dots 3}$ against the excitation point $\textrm{P}_{\textrm{0}}$ [Fig.~\ref{fig:spatial and temporal}(f)]. Overall, we still record solid amplification levels (up to $10.9 \times$ for $\textrm{P}_{\textrm{1}}$), albeit lower than those seen under sustained vibration. Then we measure the amplitude of the response after an interval equivalent to 20 excitation cycles ($A_{\mathrm{t20}}$) and we compare it against a reference amplitude $A_{\mathrm{t1,max}}$ which captures the maximum response during the excitation cycle. We see that, for $\textrm{P}_{\textrm{i}=1 \dots 3}$,  $A_{\mathrm{t20}}/A_{\mathrm{t1,max}}$ ranges from $77 \%$ to $157 \%$, indicating that the response persists even many cycles after the excitation has stopped [Fig.~\ref{fig:spatial and temporal}(h)]. 
Note that the excitation point does not experience amplitude boosting since it corresponds to an inertially constrained cantilevered site. By contrast, when the chirp spectrum contains only band-gap frequencies, the response decays rapidly in both space and time [Fig.~\ref{fig:spatial and temporal}(i)].

Since the emergence of weakly coupled resonators is obtained by microstructural retrofitting of a conventional lattice, it relies more on the inertial parameters of the microstructural correction than on the original cell design. An unexpected consequence of this fact is that our approach is not restricted to a specific geometry, but can be extended to a broad class of lattice architectures. 
To illustrate this versatility, in [Fig.~\ref{fig:generalization}] we explore numerically two such configurations, the square and triangular lattice. Figures~\ref{fig:generalization}(a) and \ref{fig:generalization}(b) show the corresponding band diagrams after the inertial anchors are introduced. Two anchoring-induced FB are highlighted by blue (OOP) and orange (IP) curves. 
The mode shapes show that the added cantilevers behave as anchors and the deformation is localized in the connecting beams. Figures~\ref{fig:generalization}(c) and \ref{fig:generalization}(d) show the spatial responses of a 5$\times$5 square and triangular lattice at the first flat-band frequency. We observe amplification factors of approximately $320 \times$ and $340 \times$, respectively, in the beams near the excitation points. In both cases, activation extends over multiple unit cells. 

In conclusion, we have described a route to realize elastic flat bands that deviates from the classical paradigm based on controlled wave interference and overcomes its limitations. Exploiting the activation of a network of resonators, we have demonstrated ability to achieve peaks of localized response while securing activation of multiple cells. The amplitude-boosting, spatial pervasiveness and temporal persistence are ideal ingredients to design lattices for vibration-based energy harvesting devices.

The authors are grateful to K. Sun and L. Jin for insightful discussions. W. M. acknowledges support of the UMN CEGE Graduate Fellowship.

\bibliographystyle{apsrev4-2}
\bibliography{Flat_band}

@article{azizi2023dynamics,
  title={Dynamics of self-dual kagome metamaterials and the emergence of fragile topology},
  author={Azizi, Pegah and Sarkar, Siddhartha and Sun, Kai and Gonella, Stefano},
  journal={Physical Review Letters},
  volume={130},
  number={15},
  pages={156101},
  year={2023},
  publisher={APS}
}

@article{samak2024direct,
  title={Direct observation of all-flat bands phononic metamaterials},
  author={Samak, Mahmoud M and Bilal, Osama R},
  journal={Physical Review Letters},
  volume={133},
  number={26},
  pages={266101},
  year={2024},
  publisher={APS}
}

@article{gonella2009interplay,
  title={Interplay between phononic bandgaps and piezoelectric microstructures for energy harvesting},
  author={Gonella, Stefano and To, Albert C and Liu, Wing Kam},
  journal={Journal of the Mechanics and Physics of Solids},
  volume={57},
  number={3},
  pages={621--633},
  year={2009},
  publisher={Elsevier}
}

@article{shen2022observing,
  title={Observing localization and delocalization of the flat-band states in an acoustic cubic lattice},
  author={Shen, Ya-Xi and Peng, Yu-Gui and Cao, Pei-Chao and Li, Jensen and Zhu, Xue-Feng},
  journal={Physical Review B},
  volume={105},
  number={10},
  pages={104102},
  year={2022},
  publisher={APS}
}

@article{riva2025creating,
  title={Creating compact localized modes for robust sound transport via singular flat band engineering},
  author={Riva, Emanuele and Bellinzoni, Federico and Braghin, Francesco},
  journal={Communications Physics},
  volume={8},
  number={1},
  pages={255},
  year={2025},
  publisher={Nature Publishing Group UK London}
}

@article{karki2023non,
  title={Non-singular and singular flat bands in tunable phononic metamaterials},
  author={Karki, Pragalv and Paulose, Jayson},
  journal={Physical Review Research},
  volume={5},
  number={2},
  pages={023036},
  year={2023},
  publisher={APS}
}

@article{sutherland1986localization,
  title={Localization of electronic wave functions due to local topology},
  author={Sutherland, Bill},
  journal={Physical Review B},
  volume={34},
  number={8},
  pages={5208},
  year={1986},
  publisher={APS}
}

@article{tasaki1994stability,
  title={Stability of ferromagnetism in the Hubbard model},
  author={Tasaki, Hal},
  journal={Physical Review Letters},
  volume={73},
  number={8},
  pages={1158},
  year={1994},
  publisher={APS}
}

@article{maksymenko2012flat,
  title={Flat-band ferromagnetism as a Pauli-correlated percolation problem},
  author={Maksymenko, M and Honecker, Andreas and Moessner, R and Richter, J and Derzhko, Oleg},
  journal={Physical Review Letters},
  volume={109},
  number={9},
  pages={096404},
  year={2012},
  publisher={APS}
}

@article{misumi2017new,
  title={New class of flat-band models on tetragonal and hexagonal lattices: Gapped versus crossing flat bands},
  author={Misumi, Tatsuhiro and Aoki, Hideo},
  journal={Physical Review B},
  volume={96},
  number={15},
  pages={155137},
  year={2017},
  publisher={APS}
}

@article{chase2024compact,
  title={Compact localized states in electric circuit flat-band lattices},
  author={Chase-Mayoral, Carys and English, LQ and Lape, Noah and Kim, Yeongjun and Lee, Sanghoon and Andreanov, Alexei and Flach, Sergej and Kevrekidis, PG},
  journal={Physical Review B},
  volume={109},
  number={7},
  pages={075430},
  year={2024},
  publisher={APS}
}

@article{zheng2014acoustic,
  title={Acoustic cloaking by a near-zero-index phononic crystal},
  author={Zheng, Li-Yang and Wu, Ying and Ni, Xu and Chen, Ze-Guo and Lu, Ming-Hui and Chen, Yan-Feng},
  journal={Applied Physics Letters},
  volume={104},
  number={16},
  year={2014},
  publisher={AIP Publishing}
}

@article{dubois2019acoustic,
  title={Acoustic flat lensing using an indefinite medium},
  author={Dubois, M and Perchoux, Julien and Vanel, AL and Tronche, Cl{\'e}ment and Achaoui, Y and Dupont, G and Bertling, K and Raki{\'c}, AD and Antonakakis, T and Enoch, S and others},
  journal={Physical Review B},
  volume={99},
  number={10},
  pages={100301},
  year={2019},
  publisher={APS}
}

@article{tang2024systematic,
  title={Systematic design and experimental realization of multiplexed acoustic double-zero-index metamaterials},
  author={Tang, Yifan and Liang, Bin and Zhu, Xuefeng and Lin, Shuyu},
  journal={Physical Review Applied},
  volume={21},
  number={3},
  pages={034020},
  year={2024},
  publisher={APS}
}

@article{leykam2013flat,
  title={Flat band states: Disorder and nonlinearity},
  author={Leykam, Daniel and Flach, Sergej and Bahat-Treidel, Omri and Desyatnikov, Anton S},
  journal={Physical Review B},
  volume={88},
  number={22},
  pages={224203},
  year={2013},
  publisher={APS}
}

@article{mukherjee2015observation,
  title={Observation of a localized flat-band state in a photonic Lieb lattice},
  author={Mukherjee, Sebabrata and Spracklen, Alexander and Choudhury, Debaditya and Goldman, Nathan and {\"O}hberg, Patrik and Andersson, Erika and Thomson, Robert R},
  journal={Physical Review Letters},
  volume={114},
  number={24},
  pages={245504},
  year={2015},
  publisher={APS}
}

@article{travkin2017compact,
  title={Compact flat band states in optically induced flatland photonic lattices},
  author={Travkin, Evgenij and Diebel, Falko and Denz, Cornelia},
  journal={Applied Physics Letters},
  volume={111},
  number={1},
  year={2017},
  publisher={AIP Publishing}
}

@article{song2023multiple,
  title={Multiple flatbands and localized states in photonic super-Kagome lattices},
  author={Song, Limin and Gao, Shenyi and Ma, Jina and Tang, Liqin and Song, Daohong and Li, Yigang and Chen, Zhigang},
  journal={Optics Letters},
  volume={48},
  number={22},
  pages={5947--5950},
  year={2023},
  publisher={Optica Publishing Group}
}

@article{azizi2024omnidirectional,
  title={Omnidirectional domain wall modes protected by fragile topological states},
  author={Azizi, Pegah and Sarkar, Siddhartha and Sun, Kai and Gonella, Stefano},
  journal={Physical Review B},
  volume={110},
  number={6},
  pages={L060102},
  year={2024},
  publisher={APS}
}

@article{fan2019elastic,
  title={Elastic higher-order topological insulator with topologically protected corner states},
  author={Fan, Haiyan and Xia, Baizhan and Tong, Liang and Zheng, Shengjie and Yu, Dejie},
  journal={Physical Review Letters},
  volume={122},
  number={20},
  pages={204301},
  year={2019},
  publisher={APS}
}

@article{Matlack-et-al_Discrete-Models_Nat-Mat_2018,
	title={Designing perturbative metamaterials from
	discrete models},
	author={Matlack, Kathryn H. and Serra-Garcia, Marc and Palermo, Antonio and Huber, Sebastian D. and Daraio, Cahira},
	journal={Nature Materials},
	volume={17},
	pages={323--328},
	year={2018},
}

@article{Riva-et-al_Nonsingular-Flat-Band_JSV_2025,
	title={Enhanced sensitivity and wave-structure interaction in nonsingular flat-band lattices with compact localized states},
	author={Riva, Emanuele and Marconi, Jacopo and Braghin, Francesco},
	journal={Journal of Sound and Vibration},
	volume={619},
	pages={119369},
	year={2025},
}

@article{kang2020topological,
  title={Topological flat bands in frustrated kagome lattice CoSn},
  author={Kang, Mingu and Fang, Shiang and Ye, Linda and Po, Hoi Chun and Denlinger, Jonathan and Jozwiak, Chris and Bostwick, Aaron and Rotenberg, Eli and Kaxiras, Efthimios and Checkelsky, Joseph G and others},
  journal={Nature Communications},
  volume={11},
  number={1},
  pages={4004},
  year={2020},
  publisher={Nature Publishing Group UK London}
}

@article{ling2026twist,
  title={Twist-Induced All-Flat-Band Higher-Order Acoustic Topological Insulator},
  author={Ling, Min-Hang and Wu, Peng and Peng, Yu-Gui and Zhu, Xue-Feng},
  journal={Advanced Materials},
  volume={38},
  number={10},
  pages={e19287},
  year={2026},
  publisher={Wiley Online Library}
}

@book{brillouin1953wave,
  title = {Wave Propagation in Periodic Structures: Electric Filters and Crystal Lattices},
  author = {Brillouin, L{\'e}on},
  edition = {2},
  publisher = {Dover Publications},
  address = {New York},
  year = {1953}
}

@article{yariv1999coupled,
  title={Coupled-resonator optical waveguide: a proposal and analysis},
  author={Yariv, Amnon and Xu, Yong and Lee, Reginald K and Scherer, Axel},
  journal={Optics Letters},
  volume={24},
  number={11},
  pages={711--713},
  year={1999},
  publisher={Optical Society of America}
}

@article{nfor2023modulational,
  title={Modulational instability and discrete localized modes in two coupled atomic chains with next-nearest-neighbor interactions},
  author={Nfor, Nkeh Oma and Yamgou{\'e}, Serge Bruno},
  journal={Journal of Nonlinear Mathematical Physics},
  volume={30},
  number={1},
  pages={71--91},
  year={2023},
  publisher={Springer}
}

@article{longhi2015tunable,
  title={Tunable dynamic Fano resonances in coupled-resonator optical waveguides},
  author={Longhi, Stefano},
  journal={Physical Review A},
  volume={91},
  number={6},
  pages={063809},
  year={2015},
  publisher={APS}
}

@article{Phani-et-al_Periodic_JASA_2006,
    author = {Phani, A. Srikantha and Woodhouse, J. and Fleck, N. A.},
    title = {Wave propagation in two-dimensional periodic lattices},
    journal = {The Journal of the Acoustical Society of America},
    volume = {119},
    number = {4},
    pages = {1995-2005},
    year = {2006},
}

@article{Celli-et-al_Culring_SMS_2017,
year = {2017},
volume = {26},
number = {3},
pages = {035001},
author = {Celli, Paolo and Gonella, Stefano and Tajeddini, Vahid and Muliana, Anastasia and Ahmed, Saad and Ounaies, Zoubeida},
title = {Wave control through soft microstructural curling: bandgap shifting, reconfigurable anisotropy and switchable chirality},
journal = {Smart Materials and Structures},
}

@article{Serra-Gracia-et-al-Quadrupole-topo-insulator_Nature_2018,
year = {2018},
volume = {555},
pages = {342-345},
author = {Serra-Garcia, Marc and Peri, Valerio and Susstrunk, Roman and Bilal, Osama R. and Larsen, Tom and Villanueva, Luis Guillermo and Huber, Sebastian D.},
title = {Observation of a phononic quadrupole topological insulator},
journal = {Nature},
}

@article{Paulose-et-al_Topological-Dislocations_Nature_2015,
year = {2015},
volume = {11},
pages = {153-156},
author = {Paulose, Jayson and Chen, Bryan Gin-ge and Vitelli, Vincenzo},
title = {Topological modes bound to dislocations in mechanical metamaterials},
journal = {Nature Physics},
}

@article{Makwana-Craster_Topological-Network_PRB_2018,
year = {2018},
volume = {98},
number = {23},
pages = {235125},
author = {Makwana, Mehul P. and Craster, Richard V.},
title = {Designing multidirectional energy splitters and topological valley supernetworks},
journal = {Physical Review B},
}

@article{deng2022observation,
  title={Observation of degenerate zero-energy topological states at disclinations in an acoustic lattice},
  author={Deng, Yuanchen and Benalcazar, Wladimir A and Chen, Ze-Guo and Oudich, Mourad and Ma, Guancong and Jing, Yun},
  journal={Physical Review Letters},
  volume={128},
  number={17},
  pages={174301},
  year={2022},
  publisher={APS}
}

@article{jo2022revealing,
  title={Revealing defect-mode-enabled energy localization mechanisms of a one-dimensional phononic crystal},
  author={Jo, Soo-Ho and Yoon, Heonjun and Shin, Yong Chang and Youn, Byeng D},
  journal={International Journal of Mechanical Sciences},
  volume={215},
  pages={106950},
  year={2022},
  publisher={Elsevier}
}

@article{sun2011nearly,
  title={Nearly flatbands with nontrivial topology},
  author={Sun, Kai and Gu, Zhengcheng and Katsura, Hosho and Das Sarma, S},
  journal={Physical Review Letters},
  volume={106},
  number={23},
  pages={236803},
  year={2011},
  publisher={APS}
}

@article{wan2023topological,
  title={Topological exact flat bands in two-dimensional materials under periodic strain},
  author={Wan, Xiaohan and Sarkar, Siddhartha and Lin, Shi-Zeng and Sun, Kai},
  journal={Physical Review Letters},
  volume={130},
  number={21},
  pages={216401},
  year={2023},
  publisher={APS}
}

@article{sarkar2025symmetry,
  title={Symmetry-based classification of exact flat bands in single and bilayer moir{\'e} systems},
  author={Sarkar, Siddhartha and Wan, Xiaohan and Lin, Shi-Zeng and Sun, Kai},
  journal={Physical Review Letters},
  volume={135},
  number={1},
  pages={016501},
  year={2025},
  publisher={APS}
}

@article{han2025all,
  title={All-angle unidirectional flat-band acoustic metasurfaces},
  author={Han, Chenglin and Fan, Shida and Zhou, Hong-Tao and He, Kuan and Jia, Yurou and Li, Changyou and Li, Hongzhu and Yang, Xiao-Dong and Chen, Li-Qun and Yang, Tianzhi and others},
  journal={Nature Communications},
  volume={16},
  number={1},
  pages={634},
  year={2025},
  publisher={Nature Publishing Group UK London}
}

@article{zhang2025observation,
  title={Observation of ultraflat bands in gapped moir{\'e} metamaterials},
  author={Zhang, Xingjian and Liu, Tingzhi and Zhang, Qicheng and Fan, Xiying and Wu, Fengcheng and Qiu, Chunyin},
  journal={Physical Review B},
  volume={111},
  number={12},
  pages={125143},
  year={2025},
  publisher={APS}
}

@article{ramos2025flat,
  title={Flat and tunable moir{\'e} phonons in twisted transition-metal dichalcogenides},
  author={Ramos-Alonso, Alejandro and Remez, Benjamin and Bennett, Daniel and Fernandes, Rafael M and Ochoa, H{\'e}ctor},
  journal={Physical Review Letters},
  volume={134},
  number={2},
  pages={026501},
  year={2025},
  publisher={APS}
}

\end{document}


\title{Supplemental Material for\\
``Giant Flat Band Amplification via Inertial Anchors''}

\maketitle

\section*{I: Theoretical analysis of the flat bands}
\label{sec:bloch}

To rationalize the formation of the flat bands, we draw an analogy between our lattice and a chain of weakly coupled local resonators. Each unit cell plays the role of one inverted-Y resonant frame in our lattice system. The dominant generalized displacement of the resonator is denoted by $q_n$, where $n$ is the unit-cell index. For an isolated resonator, the natural frequency is $\omega_0$. When these resonators are periodically arranged in the lattice, they are 
weakly coupled through the surrounding elastic network. Following the standard modal-projection and nearest-neighbor coupled-resonator approximation~\cite{brillouin1953wave}, the full elastodynamic eigenvalue problem can be reduced to an effective discrete model for the dominant resonator amplitude. Keeping only the isolated-resonator frequency and nearest-neighbor coupling, the projected mechanical eigenvalue problem is written as

\begin{equation}
    \omega^2 q_n = \omega_0^2 q_n + \kappa_2 q_{n+1} + \kappa_2 q_{n-1},
\end{equation}
where $\omega$ is the angular frequency of the collective mode and $\kappa_2$ is the effective nearest-neighbor coupling coefficient in the squared-frequency eigenvalue problem. The first term on the right-hand side represents the local resonance, while the second and third terms describe the weak coupling to the right and left neighboring resonators, respectively.

Because the system is periodic, the mode amplitude satisfies the Bloch form
\begin{equation}
    q_n = q e^{ikna},
\end{equation}
where $q$ is the amplitude within a reference unit cell, $k$ is the Bloch wave number, and $a$ is the lattice constant. Therefore,
\begin{equation}
    q_{n+1} = q e^{ik(n+1)a} = q_n e^{ika},
\end{equation}
and
\begin{equation}
    q_{n-1} = q e^{ik(n-1)a} = q_n e^{-ika}.
\end{equation}
Substituting these relations into the coupled-resonator equation gives
\begin{equation}
    \omega^2 q_n
    =
    \omega_0^2 q_n
    +
    \kappa_2 q_n e^{ika}
    +
    \kappa_2 q_n e^{-ika}.
\end{equation}
Dividing both sides by $q_n$ yields
\begin{equation}
    \omega^2(k)
    =
    \omega_0^2
    +
    \kappa_2 e^{ika}
    +
    \kappa_2 e^{-ika}.
\end{equation}
Using the identity
\begin{equation}
    e^{ika}+e^{-ika}=2\cos(ka),
\end{equation}
we obtain
\begin{equation}
    \omega^2(k) = \omega_0^2 + 2\kappa_2\cos(ka).
\end{equation}

Thus, the frequency can be written exactly as
\begin{equation}
    \omega(k)
    =
    \omega_0
    \sqrt{
    1+\frac{2\kappa_2}{\omega_0^2}\cos(ka)
    }.
\end{equation}
In the weak-coupling limit,
$|2\kappa_2/\omega_0^2|\ll 1$, this expression can be expanded to first order as
\begin{equation}
    \omega(k)
    \approx
    \omega_0
    +
    \frac{\kappa_2}{\omega_0}\cos(ka).
\end{equation}
Equivalently, by defining $\kappa=\kappa_2/(2\omega_0)$, the dispersion relation takes the tight-binding-like form~\cite{yariv1999coupled, longhi2015tunable, nfor2023modulational}
\begin{equation}
    \omega(k)
    \approx
    \omega_0+2\kappa\cos(ka).
\end{equation}

This expression shows that the local resonance frequency $\omega_0$ broadens into a narrow band when the local resonators are weakly coupled. The center frequency of the band is primarily determined by the local-resonator frequency $\omega_0$, while the bandwidth is controlled by the effective coupling strength $\kappa$. The maximum and minimum frequencies of this band are approximately
\begin{equation}
    \omega_{\max} = \omega_0 + 2|\kappa|,
\end{equation}
and
\begin{equation}
    \omega_{\min} = \omega_0 - 2|\kappa|.
\end{equation}
Thus, the bandwidth is
\begin{equation}
    \Delta \omega = \omega_{\max}-\omega_{\min} = 4|\kappa|.
\end{equation}

Based on this analogy, when the thick cantilevers act as dynamic anchors, the motion of the anchor regions is strongly suppressed. As a result, the effective coupling between neighboring resonators becomes weak, such that $|\kappa| \ll \omega_0$. In this limit, the bandwidth $\Delta\omega$ becomes very narrow and the resonant band becomes nearly dispersionless:
\begin{equation}
    \omega(k) \approx \omega_0.
\end{equation}
Therefore, the flat bands observed in the full lattice originate from weakly coupled local resonances of the inverted-Y frames, with the added cantilevers reducing the intercell coupling by acting as dynamic anchors.

\FloatBarrier

\section*{II: Effect of the thickness ratio on flat-band dispersion and amplification}
\label{sec:thickness-ratio}

The inertial contrast between the added cantilevers and the lattice beams controls the effective coupling strength $\kappa$ between neighboring resonant units and, consequently, the flatness of the resulting bands. This coupling can be tuned directly by keeping the thickness of the lattice beams fixed while varying the thickness of the cantilevers. Because the geometry of the lattice beams remains unchanged, their local resonance frequency is only weakly affected. By contrast, increasing the cantilever thickness enhances their inertia and dynamic impedance, thereby suppressing intercell coupling. The bandwidth of the targeted flat bands is therefore expected to decrease as the effective coupling strength is reduced.

\begin{figure}[H]
    \centering
    \includegraphics[width=0.93\textwidth]{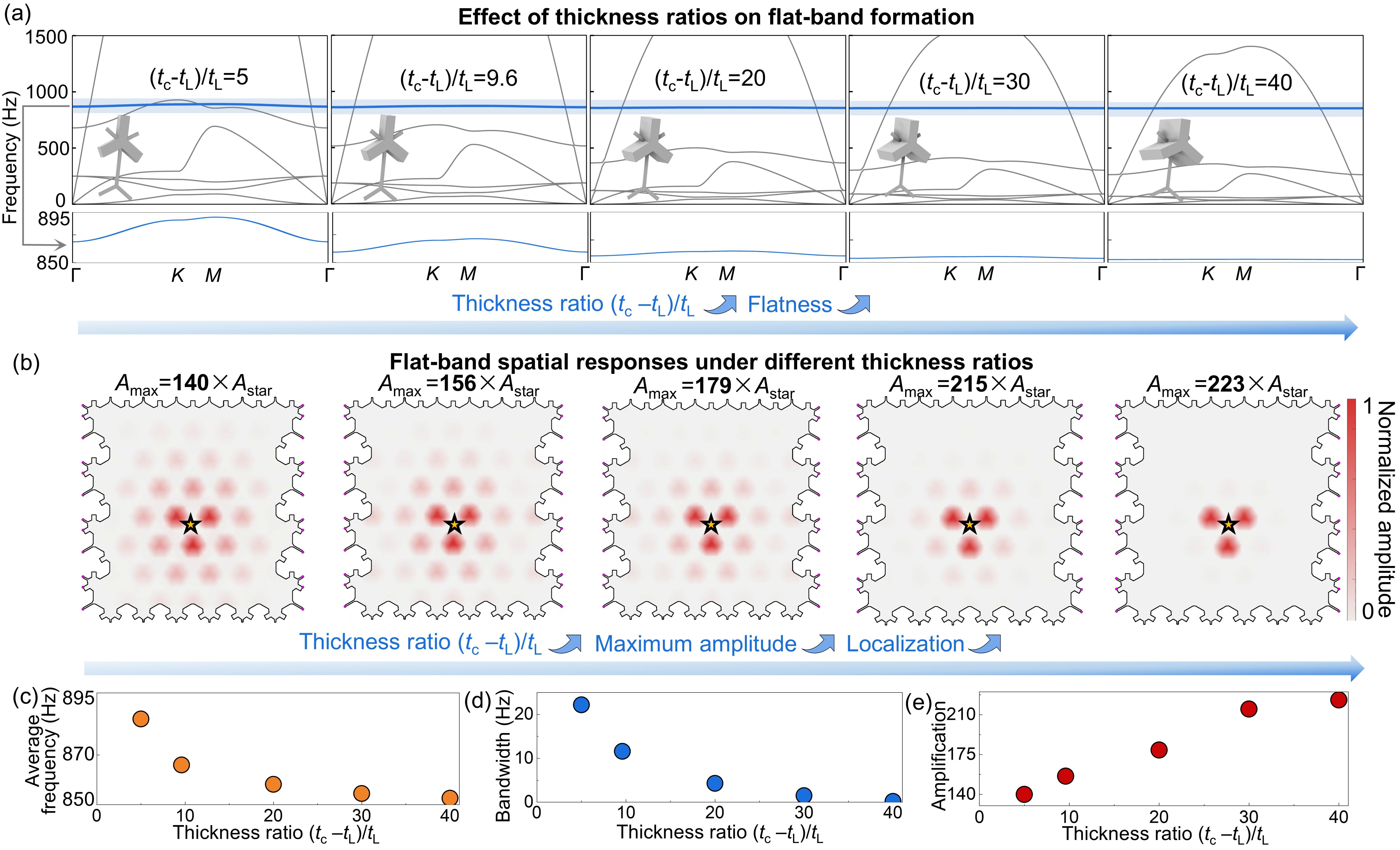}  
    \caption{Effect of thickness ratios on flat-band formation and response. 
    (a) Band diagrams of the lattice with different thickness ratios, $(t_{\mathrm{c}}-t_{\mathrm{L}})/t_{\mathrm{L}} = 5$, 9.6, 20, 30, and 40. $t_{\mathrm{c}}$ is the thickness of the added cantilever, and $t_{\mathrm{L}}$ is the thickness of the lattice. The blue curves highlight the targeted flat bands, and the enlarged views show the corresponding flat bands. Increasing ($t_{\mathrm{c}}-t_{\mathrm{L}})/t_{\mathrm{L}}$ progressively suppresses the band dispersion and leads to a nearly dispersionless flat band.
    (b) Normalized spatial responses at the corresponding flat-band frequencies for different thickness ratios. The yellow star denotes the excitation point, while the pink points along the left and right boundaries indicate the constrained locations. As ($t_{\mathrm{c}}-t_{\mathrm{L}})/t_{\mathrm{L}}$ increases, the response becomes more localized while the maximum amplification increases from $140\times$ to $223\times$.
    (c)--(e) Dependence of the average flat-band frequency, bandwidth, and peak amplification on the thickness ratios. }
    \label{fig:S1-thickness-ratio}
\end{figure}

The geometric parameters are chosen as follows. For the lattice beams, the length, out-of-plane (OOP) thickness, and in-plane (IP) width are $L_{\mathrm{L}}=50~\mathrm{mm}$, $t_{\mathrm{L}}=1~\mathrm{mm}$, and $w_{\mathrm{L}}=4~\mathrm{mm}$, respectively. For the cantilevers, the length and IP width are fixed at $L_{\mathrm{c}}=30~\mathrm{mm}$ and $w_{\mathrm{c}}=12~\mathrm{mm}$, whereas the OOP thickness is varied as $t_{\mathrm{c}}=6$, $10.6$, $21$, $31$, and $41~\mathrm{mm}$. The structure is made of aluminum, as in the main text, with Young's modulus $E=69~\mathrm{GPa}$, Poisson's ratio $\nu=0.33$, and density $\rho=3000~\mathrm{kg/m^3}$. The effect of the normalized thickness contrast, $(t_{\mathrm{c}}-t_{\mathrm{L}})/t_{\mathrm{L}}$, is summarized in [Fig.~\ref{fig:S1-thickness-ratio}].

As shown in Figs.~\ref{fig:S1-thickness-ratio}(a), (c), and (d), increasing $(t_{\mathrm{c}}-t_{\mathrm{L}})/t_{\mathrm{L}}$ substantially suppresses the dispersion of the targeted band. Specifically, the average band frequency decreases from $884$ to $852~\mathrm{Hz}$, and the bandwidth is reduced from $22.2$ to $0.2~\mathrm{Hz}$. The corresponding spatial response fields in [Fig.~\ref{fig:S1-thickness-ratio}(b)], together with the peak amplification data in [Fig.~\ref{fig:S1-thickness-ratio}(e)], reveal a simultaneous enhancement of localization and resonant amplification. As $(t_{\mathrm{c}}-t_{\mathrm{L}})/t_{\mathrm{L}}$ increases, the high-amplitude response becomes progressively confined to the vicinity of the excitation region, while the peak amplification factor increases from $140\times$ to $223\times$. These results support the dynamic-anchor mechanism: thicker cantilevers more effectively suppress coupling between neighboring resonant units, thereby producing flatter bands together with stronger localization and resonant amplification.

\FloatBarrier

\onecolumngrid

\clearpage
\twocolumngrid
\section*{III: Transmission test setup}
\label{sec:test-setup}
\begin{figure}[H]
    \centering
    \includegraphics[width=0.5\columnwidth]{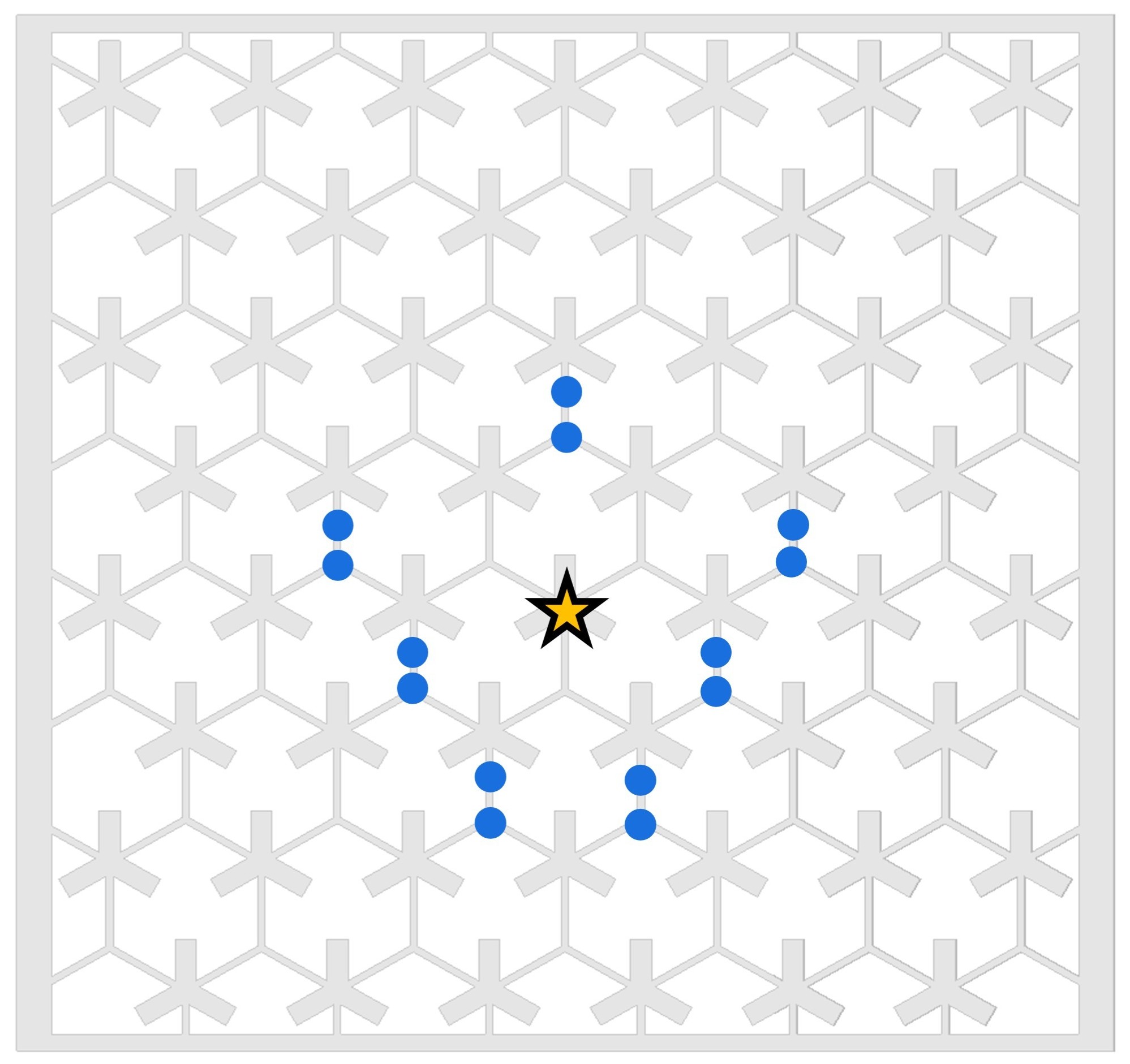}
    \caption{Spatial map of the measurement points used to construct the transmission spectrum. The yellow star denotes the excitation point located on the cantilever, while the blue circles indicate the measurement points used for calculating the transmission spectrum.}
    \label{fig:S2-transmission}
\end{figure}

For the transmission-spectrum measurement, the spatial distribution of the measurement points is shown in [Fig.~\ref{fig:S2-transmission}]. The response at the excitation point is used as the input, and the output is obtained by averaging the responses measured at the surrounding blue points. The transmission is defined as the ratio of the output response to the input response. To avoid the near-field effect associated with the local excitation, the output points are chosen from a unit cell located away from the excitation point.

\section*{IV: Band diagram and transmission of regular hexagonal lattice}
\label{sec:band-diagram}

\begin{figure}[H]
    \centering
    \includegraphics[width=0.9\columnwidth]{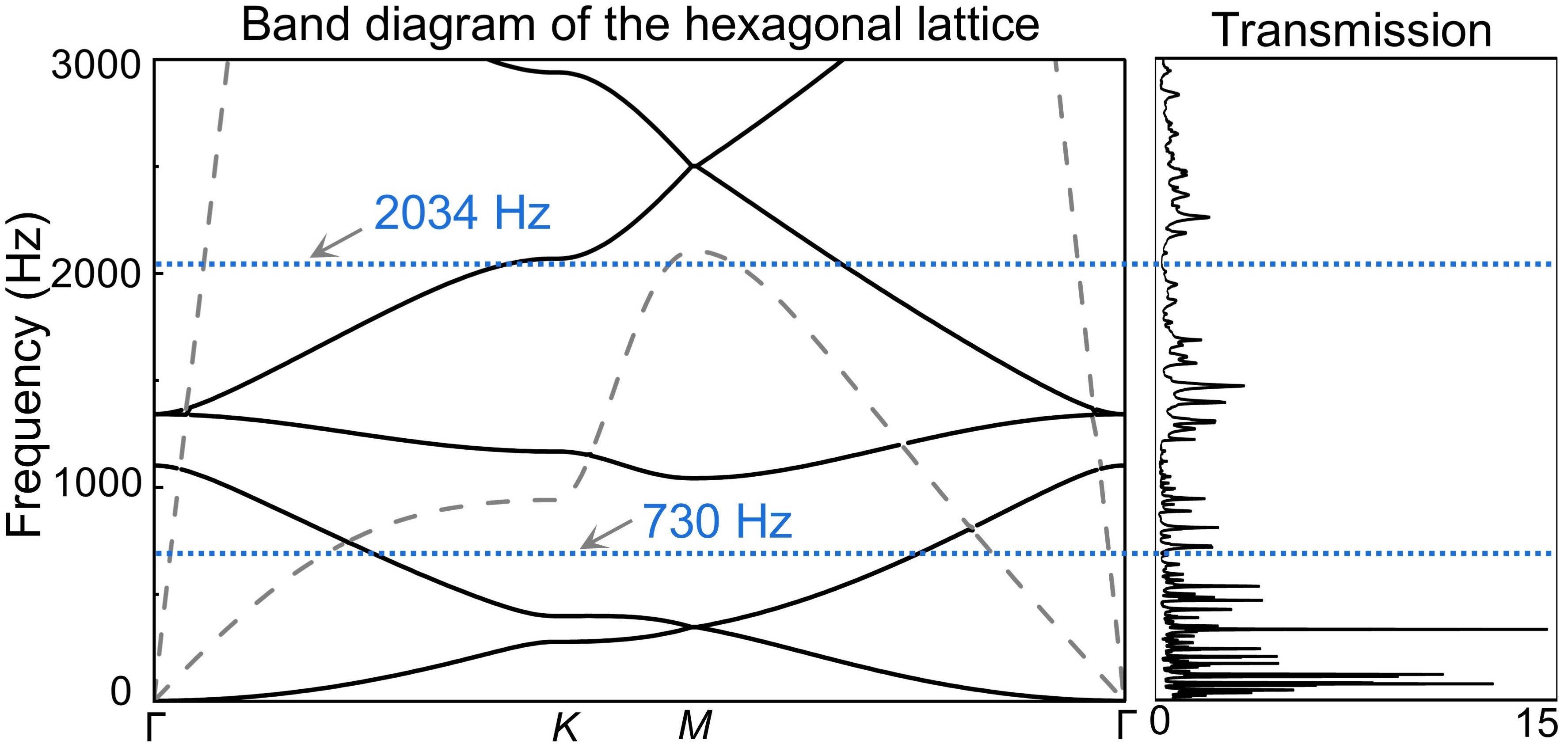}
    \caption{Numerically calculated band diagram and experimental transmission spectrum of regular hexagonal lattice. The solid lines and dashed lines represent out-of-plane (OOP) and in-plane (IP) modes, respectively.}
    \label{fig:S3-band-diagram}
\end{figure}

Figure~\ref{fig:S3-band-diagram} shows the numerically calculated band diagram and experimentally measured transmission spectrum of the regular hexagonal array used as the reference structure. The proposed flat-band lattice is obtained by retrofitting this regular lattice with inertial anchors. The flat-band frequencies of the proposed structure, 730 Hz and 2034 Hz, lie within out-of-plane pass bands of the regular hexagonal lattice, as shown in [Fig.~\ref{fig:S3-band-diagram}].

\section*{V: Geometry of the square and triangular lattice}
\label{sec:geometry}

\begin{figure}[H]
    \centering
    \includegraphics[width=0.95\columnwidth]{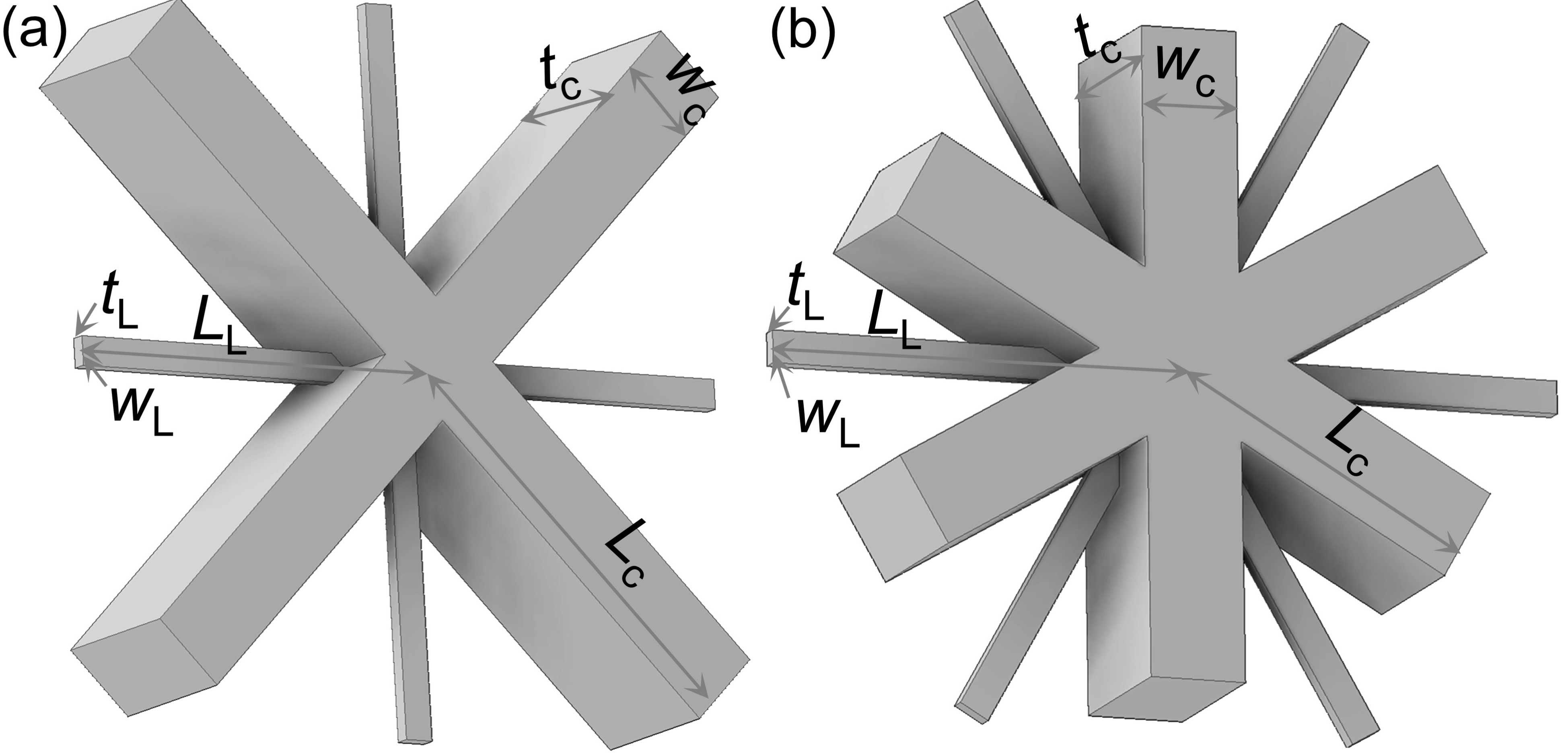}
    \caption{Geometry of the square and triangular lattice. 
    (a) Unit cell of square lattice.
    (b) Unit cell of triangular lattice.}
    \label{fig:S4-geometry}
\end{figure}

Figure~\ref{fig:S4-geometry} shows the geometries of the two lattices. Their designs are analogous to that of the hexagonal lattice considered in the main text. The number of cantilevers is determined to preserve the symmetry of the original lattice, such that the individual lattice beams act as equivalent resonators subject to identical confinement conditions. For square lattice unit cell [Fig.~\ref{fig:S4-geometry}(a)], the lattice beam length, OOP thickness and IP width $L_{\mathrm{L}}=25~\mathrm{mm}$, $t_{\mathrm{L}}=1~\mathrm{mm}$, $w_{\mathrm{L}}=2~\mathrm{mm}$, cantilever beam length, OOP thickness and IP width $L_{\mathrm{c}}=30~\mathrm{mm}$, $t_{\mathrm{c}}=11~\mathrm{mm}$, $w_{\mathrm{c}}=6~\mathrm{mm}$. For triangular lattice unit cell [Fig.~\ref{fig:S4-geometry}(b)], the lattice beam length, OOP thickness and IP width $L_{\mathrm{L}}=25~\mathrm{mm}$, $t_{\mathrm{L}}=1~\mathrm{mm}$, $w_{\mathrm{L}}=2~\mathrm{mm}$, cantilever beam length, OOP thickness and IP width $L_{\mathrm{c}}=20~\mathrm{mm}$, $t_{\mathrm{c}}=11~\mathrm{mm}$, $w_{\mathrm{c}}=6~\mathrm{mm}$. 

\bibliographystyle{apsrev4-2}
\bibliography{Flat_band}